\documentclass[aps,prl,twocolumn,superscriptaddress,showpacs]{revtex4}

\usepackage{amsmath}
\usepackage{graphics}
\usepackage{natbib}

\begin{document}

\title{Coupling and Level Repulsion in the Localized Regime: from Isolated
to Quasi-Extended Modes }
\author{K.Y. Bliokh}
\affiliation{Nonlinear Physics Centre, Research School of Physical Sciences and
Engineering, Australian National University, Canberra ACT 0200, Australia}
\affiliation{Institute of Radio Astronomy, 4 Krasnoznamyonnaya st., Kharkov 61002, Ukraine}
\author{Y.P. Bliokh}
\affiliation{Department of Physics, Technion-Israel Institute of Technology, Haifa,
32000, Israel}
\author{V. Freilikher}
\affiliation{Department of Physics, Bar-Ilan University, Ramat-Gan, 52900, Israel}
\author{A.Z. Genack}
\affiliation{Department of Physics, Queens College of the City University of New York,
Flushing, New York 11367}
\author{P. Sebbah}
\affiliation{Laboratoire de Physique de la Mati{\`{e}}re Condens{\'{e}}e, CNRS UMR6622
and Universit{\'{e}} de Nice - Sophia Antipolis, Parc Valrose, 06108, Nice
Cedex 02, France}

\begin{abstract}
We study the interaction of Anderson localized states in an open 1D random system by
varying the internal structure of the sample. As the frequencies of two states come
close, they are transformed into multiply-peaked quasi-extended modes. Level repulsion is
observed experimentally and explained within a model of coupled resonators. The spectral
and spatial evolution of the coupled modes is described in terms of the coupling
coefficient and Q-factors of resonators.

\end{abstract}

\maketitle

Transport in open disordered media can be diffusive or localized, depending
on the nature of the underlying quasimodes, which are, respectively, spread
throughout the sample or exponentially peaked within the sample, with a
typical size given by the localization length \cite{Anderson,Azbel,Sheng}.
The spatial overlap of localized modes which are close in frequency, couples
these states and leads to the formation of a series of exponential peaks
known as necklace states \cite{Mott,Lifshits,PendryPhysC,Sebbah,Wiersma}.
These states are short-lived with broadened spectral lines \cite{Wiersma,
Milner} and contribute substantially to the overall transmission in samples
much thicker than the localization length, $L\gg l_{loc}$ \cite%
{PendryPhysC,Sebbah}. Though such hybridized states are critically important
in transport and play a significant role in the localization transition,
their formation and the correlation between spatial and spectral properties
has not been explored.

In this Letter, we study the transformation of coupled Anderson localized
states in a random sample as its configuration is altered leading to the
hybridization of modes. Although level repulsion is ordinarily
associated with the diffusive regime \cite{Weidenmuller,Altshuler}, energy
level correlation and repulsion in localized 1D electron systems has been
found theoretically and numerically \cite{Gor'kov,Malyshev}. Here we present
the first experimental evidence of level repulsion of localized
electromagnetic excitations. A simple theoretical model is introduced which
explains the spectral and spatial characteristics of modes in terms of
losses within the sample and the strength of coupling between the modes.

The experiment involves a rectangular microwave waveguide opened at both
ends, which supports only a single transverse mode \cite{Sebbah}. The
waveguide is filled with a sample comprised of five 4 mm-thick blocks each
of low and high indices of refraction randomly mixed with 31 randomly
oriented 8 mm-thick binary blocks with low and high index halves. The field
inside the sample is weakly coupled to a cable which is translated along a
2-mm-wide slot cut along the waveguide in 1 mm steps. Field spectra are
measured using a vector network analyzer.

\begin{figure}[t]
\centering \scalebox{0.42} {\includegraphics{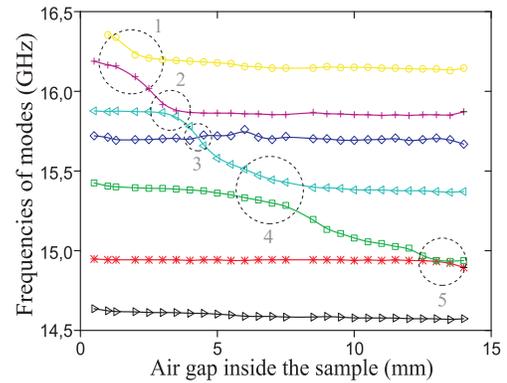}} \caption{(Color online.) Resonant
frequencies of excited localized modes vs. the driving parameter - the air gap inside the
sample. Five pair-interaction regions are circled.} \label{Fig1}
\end{figure}

Measurements are made in a sequence of configurations in which the spacing between two
segments of the sample at a depth of 60 mm is increased in steps of 0.5 mm up to a
maximum thickness of 14 mm. The position at which the air gap is introduced was chosen to
correspond to the peak of a single Anderson localized mode of the unperturbed random
sample. This allowed us to tune the frequency of the selected mode in a manner similar to
the tuning of a defect states through a band gap in a periodic structure. However, here
we deal with the Anderson localized states arising from the interference of the multiply
scattered fields in a statistically homogeneous random system. The mode frequency shifts
across the frequencies of other localized states which makes it possible to study the
coupling of two modes.

\begin{figure}[t]
\centering \scalebox{0.41}{\includegraphics{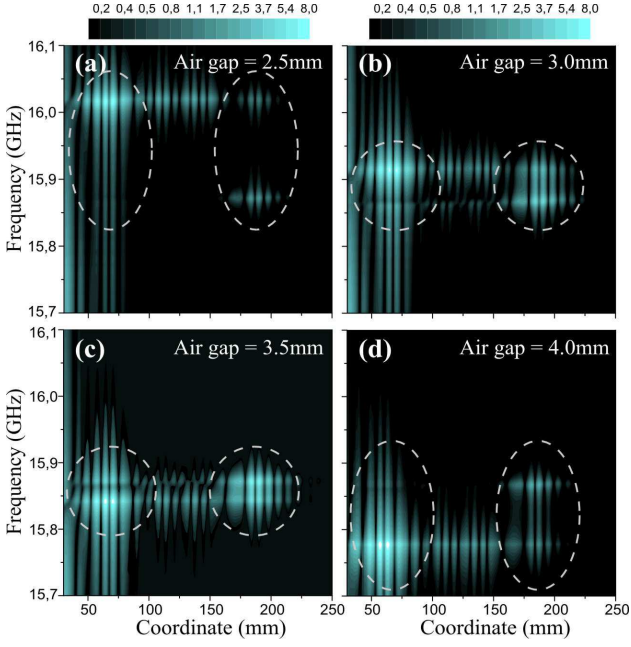}}
\caption{(Color online.) Experimentally measured normalized wave intensity
vs. frequency and coordinate inside the sample for two localized modes
corresponding to region 2 in Fig.~1 at different values of the driving
parameter. Excitations corresponding to the two effective resonators are
circled.}
\label{Fig2}
\end{figure}

The spectral positions of the localized states as functions of the air gap
introduced into the sample are plotted in Fig.~1. The frequencies of modes
may cross or exhibit an anti-crossing (level repulsion). In the latter case (regions 1,2,4,5
of Fig.~1), the coupling within the sample is accompanied by an exchange of
shape, as is seen in Fig.~2. When the frequencies of the modes are closest,
the two localized states couple into double peaked quasi-extended modes with
the same spatial intensity distribution, Fig.~2b and c. In contrast, region
3 in Fig.~1 shows a mode crossing in which the shapes are not exchanged.
This is seen in Fig.~3 which shows the driven mode passing through the broad
mode closest to the input. The two modes remain practically independent of
each other, except for the low-intensity zone (dark horizontal line in Fig.
3) at the input mode.

\begin{figure}[t]
\centering \scalebox{0.41}{\includegraphics{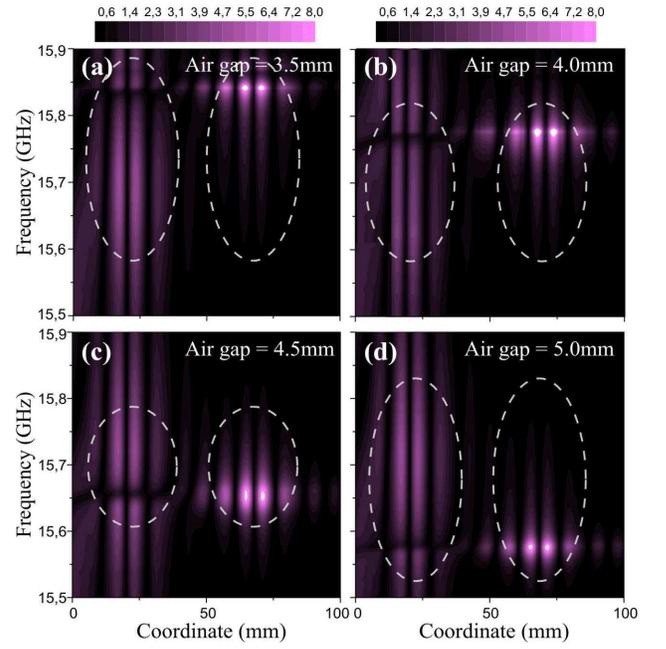}}
\caption{(Color online.) The same as in Fig.~2 for interaction region 3 in
Fig.~1.}
\label{Fig3}
\end{figure}

Resonant wave transmission through an isolated localized state in a random
sample can be described by a simple model of a wave tunneling through a
resonator with semitransparent walls \cite{Azbel,Bliokh,Rev-Mod-Phys}.
Dynamics of the field in the resonator obeys the oscillator equation with an
external force and damping, which accounts for the incident wave and the
finite Q-factor of the resonator, respectively. Extending this model to the
case of $N$ localized states which are close in frequency, we arrive at a
system of $N$ coupled oscillators with the external force acting on the
first of these:
\begin{equation}
\left\{
\begin{array}{l}
\psi _{1}^{\prime \prime }+Q_{1}^{-1}\psi _{1}^{\prime }+\left( {1-\Delta
_{1}}\right) ^{2}\psi _{1}=q_{1\,2}\psi _{2}+f_{0}e^{-i\nu \tau }~, \\
... \\
\psi _{l}^{\prime \prime }+Q_{l}^{-1}\psi _{l}^{\prime }+\left( {1-\Delta
_{l}}\right) ^{2}\psi _{l}=q_{l\,l+1}\psi _{l+1}+q_{l\,l-1}\psi _{l-1}~, \\
... \\
\psi _{N}^{\prime \prime }+Q_{N}^{-1}\psi _{N}^{\prime }+\left( {1-\Delta
_{N}}\right) ^{2}\psi _{N}=q_{N\,N-1}\psi _{N-1}~.%
\end{array}
\right.  \label{1}
\end{equation}
Here $\psi _{i}\left( \tau \right) $ is the field in the $i$th resonator, $%
\tau =\omega _{0}t$ is the dimensionless time ($\omega _{0}$ is a
characteristic central frequency of the problem), $1-\Delta _{i}$ ($|\Delta
_{i}|\ll 1)$ is the dimensionless eigenfrequency of $i$th resonator, $%
Q_{i}\gg 1$ is the Q-factor describing the losses of the energy in the $i$th
resonator, $q_{i\,i+1}=q_{i+1\,i}\ll 1$ is the coupling coefficient of $i$th
and $(i+1)$th resonators due to the spatial overlap of their modes; $f_{0}$
and $\nu $, ($|\nu -1|\ll 1)$ are the amplitude and frequency of the
external field exciting the first resonator. The Q-factors can be written as
\cite{Yu. Bliokh}:
\begin{equation}
Q_{i}^{-1}=\Gamma _{i}~(1<i<N)~,~~Q_{1,N}^{-1}=\Gamma _{1,N}+\frac{
v_{g}T_{in,out}}{2l\omega _{0}}~,  \label{2}
\end{equation}
where $\Gamma _{i}$ is the dissipation rate in the $i$th resonator, $%
T_{in,out}$ are the transmission coefficient of the input and output of the
system, $v_{g}$ is the wave group velocity inside the resonator cavity, and $%
l$ is the cavity length. The last term in Eqs.~(2) accounts for the energy
leakage through the outermost resonators.

\begin{figure}[t]
\centering \scalebox{0.42}{\includegraphics{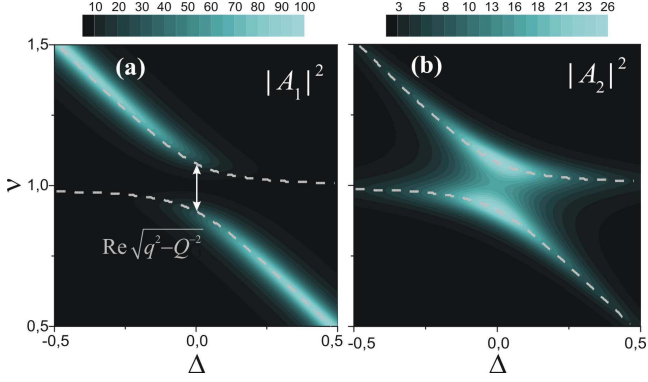}} \caption{(Color online.) Field
amplitudes $|A_{1,2}|^{2}$ in the two resonators as functions of the incident field
frequency $\protect\nu $ and
detuning $\Delta $ between the resonators. The underlying frequencies $%
\protect\delta\protect\nu _{\mathrm{res}}^{\pm }$ are depicted by the dashed
lines. Parameters are: $f_{0}=1$, $q=0.2$, and $Q^{-1}=0.1$ ($Q^{-1}<q $
regime).}
\label{Fig4}
\end{figure}

To establish the correspondence between the model Eq.~(1) and localized
states in a random sample, we assume, following \cite{Bliokh,Rev-Mod-Phys},
that $\psi _{i}$ represents the peak field of the $i$th localized state, $%
q_{i\,i+1}\psi _{i}$ is the amplitude of the field penetrating the adjacent
cavity, and $f_{0}$ is the amplitude of the incident wave, $\psi _{0}$,
penetrating the first localization cavity. Since for long enough system the
localization length is the only disorder-induced spatial scale in the
problem, we assume
\begin{eqnarray}
q_{i\,i+1} &\simeq &\exp \left( {-d_{i\,i+1}/l_{loc}}\right) ,~f_{0}\simeq
\psi _{0}\exp \left( {-d_{in}/l_{loc}}\right) ,  \notag  \label{3} \\
T_{in,out} &\simeq &\exp \left( {-2d_{in,out}/l_{loc}}\right) ~,~~l\sim
l_{loc}~.
\end{eqnarray}%
Here $d_{i\,i+1}=\left\vert {X_{i+1}-X_{i}}\right\vert $ is the distance
between adjacent states at coordinates $X_{i+1}$ and $X_{i}$, whereas $%
d_{in}=X_{1}$ and $d_{out}=L-X_{N}$ are the distances from the edge
localized states to the corresponding ends of the sample. The localization
length is determined through the average value of the total transmission
coefficient $T$ as $l_{loc}^{-1}=-<\ln T(L)>/2L$ [3]. The deterministic
equations (1)--(3) provide an effective description of coupled modes in 1D
random system.

Substituting $\psi _{i}=A_{i}\exp \left( {\ -i\nu \tau }\right) $, the set
of equations (1) is reduced to an algebraic equation $\hat{H}\vec{\Psi}=\vec{%
F}$ with
\begin{equation}
\hat{H}=\left( {\begin{array}{*{20}c} {C_1 } & { - q_{1,2} } & {...} & 0 & 0
\\ { - q_{1,2} } & {...} & { - q_{l - 1,l} } & 0 & 0 \\ {...} & { - q_{l -
1,l} } & {C_l } & { - q_{l,l + 1} } & {...} \\ 0 & 0 & { - q_{l,l + 1} } &
{...} & { - q_{N - 1,N} } \\ 0 & 0 & {...} & { - q_{N - 1,N} } & {C_N } \\
\end{array}}\right) ,  \label{4}
\end{equation}
$\vec{\Psi}=\left( A_{1},...,A_{N}\right) ^{T}$, and $\vec{F}=f_{0}\left(
1,0,...,0\right) ^{T}$, where $C_{i}=\left( {1-\Delta _{i}}\right) ^{2}-\nu
^{2}-i\nu Q_{i}^{-1}\simeq 2\left( {1-\Delta _{i}-\nu }\right) -iQ_{i}^{-1}$%
. The homogeneous equation $\hat{H}\vec{\Psi}=0$ determines a set of
 eigenmodes, with eigenfrequencies being the
eigenvalues of the matrix (4).

For the sake of simplicity, we consider the case of two interacting modes, $%
N=2$, and assume that $Q_{1}=Q_{2}\equiv Q$ and $q_{1\,2}\equiv q$. Then,
the complex eigenfrequencies $\nu ^{\pm }=1+\delta \nu ^{\pm }$ are given by
\begin{equation}
\delta \nu ^{\pm }=-\frac{\Delta _{1}+\Delta _{2}}{2}-i\frac{Q^{-1}}{2}\pm
\frac{1}{2}\sqrt{(\Delta _{1}-\Delta _{2})^{2}+q^{2},}  \label{res-1}
\end{equation}
This equation describes level anti-crossing  and the
coupling of isolated resonators to form collective eigenmodes. The minimal
frequency gap $q$ is achieved at resonance, $\Delta_1 =\Delta_2$. Away from
resonance, $|\Delta_1-\Delta_2|\gg q$, the eigenmodes tend to the modes of
isolated resonators. The shapes of the modes are exchanged when passing
through the resonance, i.e., $+$ ($-$) eigenmodes correspond to the first
(second) resonator at $\Delta _{1}\ll \Delta _{2}$ and to the second (first)
resonator when $\Delta _{1}\gg \Delta _{2}$. It is important to note that
level repulsion of electromagnetic modes arises in a finite system in the regime of strong
localization.

\begin{figure}[t]
\centering \scalebox{0.42}{\includegraphics{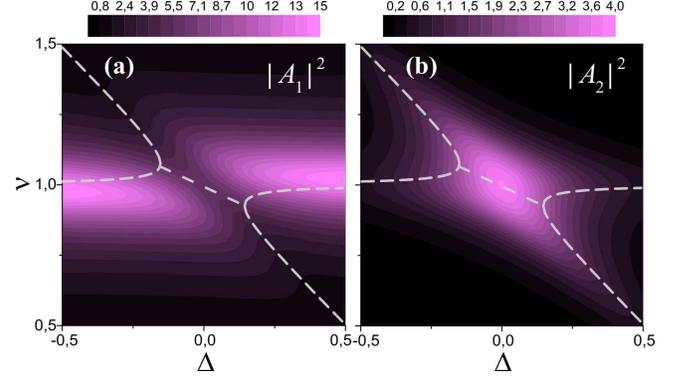}} \caption{(Color online.) The same
as in Fig.~4 for the parameters $q=0.2$ and $Q^{-1}=0.25$ ($Q^{-1}>q$ regime).}
\label{Fig5}
\end{figure}

If the system is excited by a monochromatic wave with real frequency $\nu
=1+\delta \nu $, as in the experiment, the complex amplitudes $A_{1,2}$ of
the two resonators can be obtained from $\hat{H}\vec{\Psi}=\vec{F}$, which
yields
\begin{eqnarray}
A_{1} &=&-\frac{{\left[ {2\left( {\Delta _{2}+\delta \nu }\right)
+iQ_{2}^{-1}}\right] f_{0}}}{D}~,~~A_{2}=\frac{{qf_{0}}}{D},  \label{6} \\
D &=&\left[ {2\left( {\Delta _{1}+\delta \nu }\right) +iQ_{1}^{-1}}\right] %
\left[ {2\left( {\Delta _{2}+\delta \nu }\right) +iQ_{2}^{-1}}\right] -q^{2}
\notag
\end{eqnarray}
The behavior of $|A_{1,2}|^{2}$ is essentially determined by the denominator
$|D|^{2}$, which is at minimum at frequencies
\begin{equation}
\delta \nu _{\mathrm{res}}^{\pm }=-\frac{(\Delta _{1}+\Delta _{2})}{2}\pm
\frac{1}{2}\mathrm{Re}\sqrt{(\Delta _{1}-\Delta _{2})^{2}+q^{2}-Q^{-2}}.
\label{7}
\end{equation}
The amplitudes $A_{1,2}$ and frequencies $\delta \nu _{\mathrm{res}}^{\pm }$
characterize the resonant excitation of the system by an external source.
Note that Eq.~(7) coincide with Eq. (5) only in the lossless case $Q^{-1}=0$%
. Otherwise, there are two different regimes of the excitation of coupled
resonators, determined by the ratio between losses $Q^{-1}$ and coupling $q$%
. If losses are small, $Q^{-1}<q$, two branches $\delta \nu_{\mathrm{res}%
}^{\pm }$ demonstrate anti-crossing with a frequency gap $\sqrt{q^{2}-Q^{-2}}
$, Fig.~4. If losses are greater than the coupling strength, so that $%
Q^{-1}>q$, the frequencies $\delta \nu ^{\pm }$ merge in the interaction
region $(\Delta _{1}-\Delta _{2})^{2}\leq |q^{2}-Q^{-2}|$, Fig.~5.

The amplitudes (6) for $Q^{-1}<q$ and $Q^{-1}>q$ are shown as functions of
the frequency $\nu $ and detuning $\left( \Delta _{1}-\Delta _{2}\right) $
in Figs.~4 and 5, respectively. The main features observed experimentally
are reproduced. Different values of $\Delta $ correspond to different frames
of Figs. 2 and 3. To facilitate comparison with the experimental results in
Figs.~2 and 3, the second, output resonator is driven in Fig.~4 ($\Delta
_{1}\equiv 0$, $\Delta _{2}\equiv \Delta $), while the first resonator is
driven in Fig.~5 ($\Delta _{1}\equiv \Delta $, $\Delta _{2}\equiv 0)$). It
is seen in Figs. 2 and 4 that for $Q^{-1}<q$ fields in both resonators in
the interaction regime (Figs. 2b and c) exhibit double-peaked spectra (level
repulsion). Collective excitation of two resonators signifies the formation
of quasi-extended necklace states. Remarkably, away from the resonance
(Figs. 2a and d), the first resonator is effectively excited at one of the
resonant frequencies, close to $\delta \nu =-\Delta _{1}$ (Fig.~4a), while
the second resonator is equally excited at both resonant frequencies $\delta
\nu _{\mathrm{res}}^{\pm }\simeq \Delta _{1,2}$ (Fig.~4b). In the regime $%
Q^{-1}>q$, both Figs. 3 and 5 show that the second resonator is excited with
a single peak in the spectrum (Fig.~5b), while two peaks separated by a dark
area driven with the frequency of the second resonator are seen in the first
resonator (Fig.~5a) \cite{Note}.

\begin{figure}[t]
\centering \scalebox{0.40}{\includegraphics{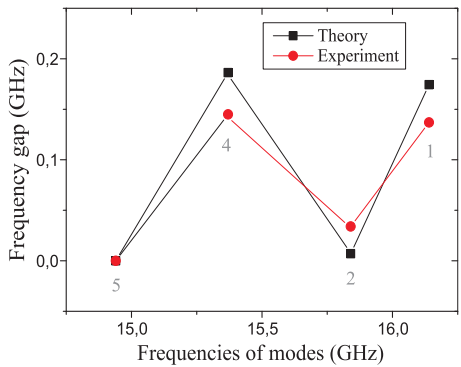}}
\caption{(Color online.) Experimentally measured and theoretically
calculated ($\mathrm{Re}\protect\sqrt{q^2-Q^{-2}}$) minimal frequency gaps
for pairs of interacting modes 1,2,4,5 presented in Fig.~1.}
\label{Fig6}
\end{figure}

The measured and calculated values of the frequency gap between coupled
modes are presented in Fig.~6. The parameters of the system are:
\begin{equation}
\frac{\omega _{0}}{2\pi }\simeq 15.5\mathrm{GHz},~l_{loc}\simeq 12\mathrm{mm}%
,~\omega _{0}\Gamma \simeq 7\times 10^{7}\mathrm{s}^{-1},  \label{8}
\end{equation}%
and $v_{g}\simeq c/2.4$, whereas the positions of the localized modes
interacting in the regions 1--5 (Fig.~1) are equal, respectively, to
\begin{eqnarray}
X_{1} &\simeq &~64:~64:~~7:~64:~64~\mathrm{mm}~,  \notag \\
X_{2} &\simeq &117:192:64:128:235~\mathrm{mm}~.  \label{9}
\end{eqnarray}%
Substituting values (8) and (9) into Eqs.~(2) and (3) yields $Q_{1,2}^{-1}$
and $q$. We calculated the minimum frequency difference for interacting
pairs 1,2,4,5 (for which $Q_{1}^{-1}\sim Q_{2}^{-1}$ \cite{Note}) as $%
\mathrm{Re}\sqrt{q^{2}-Q^{-2}}$ with $Q^{-1}=(Q_{1}^{-1}+Q_{2}^{-1})/2$, and
compared it to measured values of the gap, Fig.~1. Good agreement between
the experiment and model is seen in Fig.~6.

In conclusion, we have observed level repulsion in the localization
regime and have shown that it reflects the coupling of localization centers.
The occurrence of anti-crossing or crossing of quasimodes as a sample
configuration changes depends upon the ratio of the coupling strength
between localized states and loss. These factors
determine the statistics of level spacings and widths  and thus the diffusive or localized regime of wave propagation.

We thank B. Hu, J. Klosner, H. Rose, and Z. Ozimkowski for their
contributions to the implementation of the waveguide assembly, and L. Pastur for illuminating discussion. This research
was sponsored by the Linkage International Grant of the Australian Research
Council, National Science Foundation (DMR-0538350), PSC-CUNY, the Centre
National de la Recherche Scientifique (PICS $\#2531$ and PEPS07-20), and the
Groupement de Recherches IMCODE.


\begin{thebibliography}{99}
\bibitem{Note} It should be noted that two
localized modes of Fig.~3 have significantly different Q-factors, $%
Q_{1}^{-1}\gg Q_{2}^{-1}$. This diminishes the coupling between the modes
since their frequencies are distant in the complex plane. Nonetheless, our
simplified model with $Q_{1}=Q_{2}$ describes quantitatively all the main
features of the coupling in this case.
\end{thebibliography}
\end{document}